\documentclass{jaa}
\usepackage{natbib}
\usepackage{aas_macros}
\usepackage[unicode=true,pdfusetitle,
 bookmarks=true,bookmarksnumbered=false,bookmarksopen=false,
 breaklinks=false,pdfborder={0 0 0},pdfborderstyle={},backref=false,citecolor=violet,colorlinks=true]
 {hyperref}
\usepackage{amsmath}
\usepackage{graphicx}
\usepackage{xcolor}
\usepackage{ulem}
\usepackage{bm}
\usepackage{soul}

\newcommand{\EnsAve}[1]{\big\langle #1 \big\rangle}
\newcommand{\vecHat}[1]{\hat{\mathbf{#1}}}

\begin{document}\sloppy

\title{Probing Cosmology beyond $\Lambda$CDM using the SKA}


\author{Shamik Ghosh\textsuperscript{1,2}, Pankaj Jain\textsuperscript{3}, Rahul Kothari\textsuperscript{4,*},  Mohit Panwar\textsuperscript{3},  Gurmeet Singh\textsuperscript{3}, Prabhakar Tiwari\textsuperscript{5}}
\affilOne{\textsuperscript{1}CAS Key Laboratory for Researches in Galaxies and Cosmology, Department of 
 Astronomy, University of Science and Technology of China, Chinese Academy
of Sciences, Hefei, Anhui 230026, China\\}
\affilTwo{\textsuperscript{2}School of Astronomy and Space Science, University of Science and Technology of China, Hefei, 230026, China\\}
\affilThree{\textsuperscript{3}Department of Physics, Indian Institute of Technology, Kanpur-208016, India.\\}
\affilFour{\textsuperscript{4}Department of Physics \& Astronomy, University of the Western Cape, Cape Town 7535, South Africa.\\}
\affilFive{\textsuperscript{5}National Astronomical Observatories, Chinese Academy of Science, Beijing  100101, P.R.China.\\}


\twocolumn[{

\maketitle

\corres{quantummechanicskothari@gmail.com}

\msinfo{1 January 2022}{1 January 2022}

\begin{abstract}
The cosmological principle states that the Universe is statistically homogeneous and isotropic at large distance scales. There currently exist many observations which indicate a departure from this principle. It has been shown that many of these observations can be explained by invoking superhorizon cosmological perturbations and may be consistent with the Big Bang paradigm. Remarkably, these modes simultaneously explain the observed Hubble tension, i.e., the discrepancy between the direct and indirect measurements of the Hubble parameter. We propose several tests of the cosmological principle using SKA. In particular, we can reliably extract the signal of dipole anisotropy in the distribution of radio galaxies. The superhorizon perturbations also predict a significant redshift dependence of the dipole signal which can be nicely tested by the study of signals of reionization and the dark ages using SKA. We also propose to study the alignment of radio galaxy axes as well as their integrated polarization vectors over distance scales ranging from a few Mpc to Gpc. We discuss data analysis techniques that can reliably extract these signals from data.
\end{abstract}


}]

\section{Introduction \label{sec:Intro}}


Current observations support an expanding universe. If we extrapolate this back in time,  we can infer that the Universe started from a very hot and dense state. This event, known as \textit{Big Bang}, marked the origin of the Universe in a very high temperature state.  

In order to make the problem of expansion dynamics tractable, we assume that the Universe is {spatially} isotropic and homogeneous. This assumption is also known as Cosmological Principle (hereafter CP) \citep{Kolb1994,Einstein:1917,Aluri:2022hzs}. {It turns out that} Hubble's law is a direct consequence of CP \citep{coles2003}. Furthermore, it can be shown that the most general spacetime metric that describes a universe following CP is the FLRW metric \citep{weinberg1972,coles2003}.  It is also important to mention that CP is an independent {assumption} and does not follow from symmetries of the Einstein's Equations.

The FLRW metric describes {a Universe with a smooth background having an exact isotropic and homogeneous matter distribution.} 
{But observationally}, the Universe also possesses structure in the form of stars, galaxies, etc. These structures arise due to {curvature} perturbations which are seeded during {the epoch of exponential expansion} called \textit{inflation}. The resulting cosmological model, including dark matter and dark energy is called $\Lambda$CDM.

Although, these perturbations aren't isotropic and homogeneous \textit{per se}, they satisfy these properties in a statistical sense. For example, in the cosmic frame of rest, the matter density is expected to be the same at all points provided we average over a sufficiently large distance scale. The precise value of this distance scale is still not clear but is expected to be of order 100 Mpc (see, for example \cite{Kim:2021osl}).

It has been speculated that {during an epoch}, before inflation {ensued}, the Universe may be described by a complicated metric whose nature is currently poorly understood. However, it quickly evolves to the isotropic and homogeneous FLRW metric during inflation, perhaps within the first e-fold. {\cite{Wald:1983} for the first time, gave an explicit demonstration for Bianchi Universes. {Some other results also exist for inhomogeneous metric \citep{Stein-Schabes:1986lic,Jensen:1986vs}}.  We may speculate that the idea generalizes to a larger class of metrics\footnote{{There are exceptions to these results as well \citep{sato_1988}.}}.}
The Big Bang paradigm is therefore consistent with an early anisotropic and/or inhomogeneous phase of the Universe. 
Given the existence of such a phase, it is clearly important to ask whether it has any observational consequences.

Observationally, 
the Universe is found to be consistent with CP to a good approximation. 
But currently there exist many observations in CMB and large scale structures (LSS henceforth) which appear to violate CP \citep{Shamik:2016}. We review these anomalies later in  \S\ref{sec:anomalies} {For an expansive review, see \citet{Aluri:2022hzs}.} There exist many theoretical attempts to explain these observations. It has been suggested that \textit{superhorizon modes}, i.e., perturbations of wavelengths larger than the horizon size \citep{Grishchuk:1978AZh,Grishchuk:1978SvA}, may explain some of these observations \citep{Gordon:2005,Erickcek:2008a,Erickcek:2008b,Ghosh:2014,Das:2021JCAP,Tiwari:2022HT}. {Additionally, these can  account for low-$\ell$ alignments \citep{Gao:2009vi}, though these can't extenuate the present accelerated expansion of the Universe \citep{Hirata:2005ei,Flanagan:2005dk}.} It is assumed that such large wavelength modes are aligned with one another and hence do not obey CP. An intriguing possibility is that such modes might originate during an anisotropic and/or inhomogeneous pre-inflationary phase of the Universe \citep{Aluri:2012Pre-Inf,Rath:2013}. Hence, despite being in violation with CP, they would be consistent with the Big Bang paradigm.

\subsection{Mathematical Formulation and Ramifications}
\begin{figure}
\begin{center}
    \includegraphics[scale=0.5]{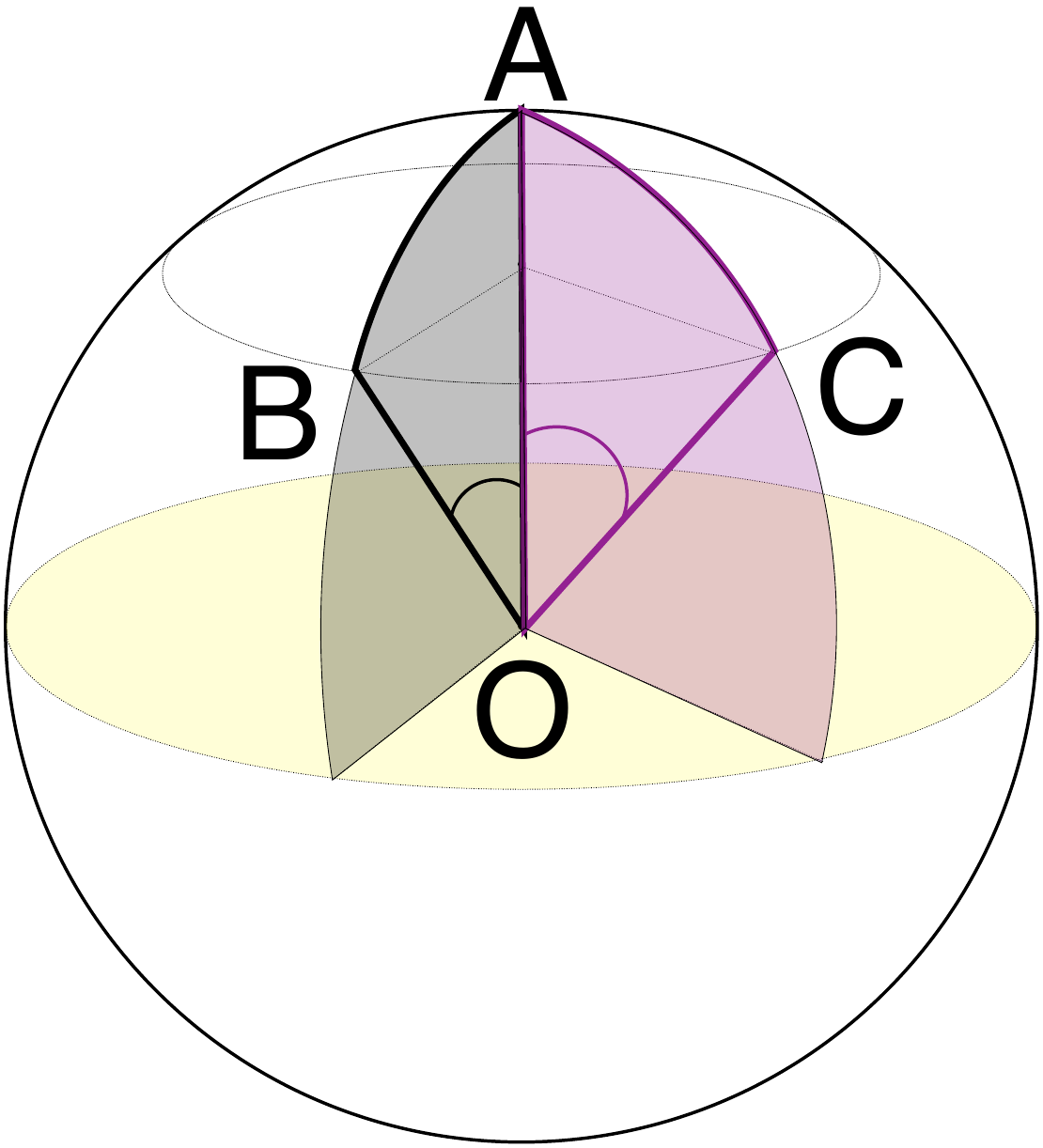}
\end{center}
    \caption{Illustration of statistical isotropy. In this Figure, $A$, $B$ and $C$ are given points on the spherical surface such that $\angle AOB=\angle AOC$.}
    \label{fig:StatIso}
\end{figure}

In order to relate theory with observations, we seek  ensemble averages of the fields under consideration. Ergodicity hypothesis \citep{ellis_maartens_maccallum_2012} allows us to relate this ensemble averaging to the space averaging. It is known that for the gaussian random fields, all the statistical information is contained in the 2 point correlation functions (2PCF). However, in the presence of non-gaussianities, we need higher order correlators like bispectrum (3PCF) or trispectrum (4PCF), etc., in order to extract optimal cosmological information. CP dictates that the  nPCF only be a function of distances between the points $\mathbf{x}_i\equiv (z_i,\mathbf{n}_i)$. Thus
\begin{equation}
   \big\langle \rho(\mathbf{x}_1)\rho(\mathbf{x}_2)\ldots \rho(\mathbf{x}_n) \big\rangle = f(x_{12},x_{13},\ldots,x_{ij},\ldots),\label{eq:CosmoPrin}
\end{equation}
where $x_{ij}=|\mathbf{x}_i-\mathbf{x}_j|=x_{ji}$ and $i\ne j$.
Clearly, this makes this nPCF invariant under arbitrary translations and rotations. {The} condition \eqref{eq:CosmoPrin} for 2PCF in case of a 2D field, e.g., CMB temperature, takes the usual form
\begin{equation}
    \EnsAve{T(\mathbf{x}_1)T(\mathbf{x}_2)}\equiv\EnsAve{T(\vecHat{m})T(\vecHat{n})}=f(\vecHat{m}\cdot\vecHat{n}),\label{eq:2PCFCMB}
\end{equation}
with $\mathbf{x}_1\equiv (z_*,\vecHat{n})$ and $\mathbf{x}_2\equiv (z_*,\vecHat{m})$, {$z_*$ being the redshift to decoupling}.
Eq. \eqref{eq:2PCFCMB} is the familiar result for the 2PCF, which dictates that the temperature correlation depends only upon the angle between the locations. This is illustrated in Figure \ref{fig:StatIso}, where three points $A$, $B$ and $C$ are chosen in a manner such that $\angle AOC=\angle AOB$. Thus we must have $\EnsAve{T(\mathbf{A})T(\mathbf{C})}=\EnsAve{T(\mathbf{A})T(\mathbf{B})}$, since $\mathbf{A\cdot C}=\mathbf{A\cdot B}$.

\section{Observations at tension with $\Lambda$CDM}
\label{sec:anomalies}
Our observations in the past two decades have firmly planted the inflationary $\Lambda$CDM cosmology as the standard paradigm. A vast set of observables from CMB to LSS broadly agree with $\Lambda$CDM predictions. Despite the successes of $\Lambda$CDM, we have a growing set of observations that are at tension with our expectations from $\Lambda$CDM. We will summarise some of the observed tensions, in the context of the model discussed in this paper. {See \citet{Perivolaropoulos:2021jda} for a review.}

\subsection{Observed violations of Statistical Isotropy}

As we discussed before, CP implies statistical isotropy and homogeneity. Due to our fixed vantage point, it is not possible to directly test statistical homogeneity. However, we can test statistical isotropy. Various observational tests, performed on different cosmological datasets, amply attest statistical isotropy violations. Some of these are reviewed in \citet{Shamik:2016}. 

\subsubsection{The kinematic dipole:\label{sec:KineDipo}}
As explained in the \S\ref{sec:Intro}, CP is valid only in the cosmic frame of rest. We as observers are \textit{not} stationary with respect to this frame on account of the motion of Earth, the Sun, and the Milky Way. This gives rise to an effective peculiar velocity to our observation frame. This peculiar velocity results in a Doppler boost of the CMB temperature fluctuation, further culminating in a kinematic dipole in the CMB temperature fluctuations. Interpreting the CMB dipole to be of kinematic origin \citep{Planck:2014} leads to the peculiar velocity of our local frame to be $384 \pm 78 \text{ km s}^{-1}$.  

The peculiar velocity $\mathbf v$ of our observation is expected to give rise to a dipole in the observed number count of sources. The local motion would cause a Doppler and aberration effects, both of which contribute to a dipole in the observed number counts \citep{Ellis:1984}. For sources with flux following a power law relation in frequency: $S \propto \nu^{-\alpha}$, 
and with differential number count $N(S, \mathbf{\hat {n}}) = S^{-1-x}$, the expected dipole is given by:
\begin{equation}
    \mathbf D = \left[2 + x(1+\alpha) \right] \mathbf v / c,\label{eq:DipNoSuperHori}
\end{equation}
where $c$ is the speed of light, $\alpha$ {is the frequency scaling spectral index}, and $(1+x)$ {is  the slope of $\ln N$ v/s $-\ln S$ plot}. We can use the estimates of our peculiar velocity from the CMB and use it to predict the estimated dipole in the large-scale structure data. Assuming $\alpha \approx 0.75$ and $x \approx 1$, we find the expected dipole $D_{\rm th} \sim 0.005$. Measurements of the dipole in LSS surveys at $z \sim 1$ have all yielded results that are consistent with CMB direction but the magnitude is found to be double or more of the predicted value. In Table \ref{tab:lss_dipole}, we list the measured value of dipole in the NVSS, NVSS+WENSS, NVSS+SUMSS and CatWISE catalogs.  The dipole measured in the LSS has a much larger magnitude than expected from CMB measurements but is consistent with the CMB dipole direction. The deviation is found to be at $4.9 \sigma$ in the CatWISE data \citep{Secrest_2021}. 

{
We point out that the assumed power law dependence of number counts on $S$ is not strictly valid \citep{Tiwari:2014ni}. This leads to a difference in the dipole in number counts and in sky brightness. It also introduces a dipole in the mean flux per source. Hence this provides a nontrivial test of whether the dipole is indeed of kinematic origin. This idea has been generalised in \cite{Nadolny_2021} who develop a method to  extract kinematic dipole independently from an intrinsic dipole. }

\begin{table}[t]
    \centering
    \begin{tabular}{lcc} 
    \hline \hline \\
        Authors  & $|\mathbf{D}|$ ($\times 10^{-2}$) & ($l, b$) \\[2ex]
    \hline \\[-1ex]
        \citet{Singal:2011} & $1.8\pm 0.3$ & ($239^\circ, 44^\circ$) \\
        \citet{Rubart:2013} & $1.6\pm 0.6$ & ($241^\circ, 39^\circ$) \\
        \citet{Tiwari:2014ni} & $1.25\pm 0.40$ & ($261^\circ, 37^\circ$) \\
        \citet{Tiwari:2016adi} & $0.9 \pm 0.4$ & ($246^\circ, 38^\circ$)\\
        \citet{Colin:2017} & $1.6 \pm 0.2$& ($241^\circ, 28^\circ$)\\
        \citet{Secrest_2021} & $1.5 $& ($238^\circ, 29^\circ$)\\
      \hline  \hline
    \end{tabular}
    \caption{Results for the dipole in LSS exceed the expected value of $5 \times 10^{-3}$.}
    \label{tab:lss_dipole}
\end{table}

\subsubsection{Alignment of quadrupole $(\ell=2)$ and octupole $(\ell=3)$:} Both  $\ell=2,3$ CMB multipoles are aligned with preferred direction pointing roughly along the CMB dipole \citep{deOliveira-Costa:2003utu}. Physically, both of these multipoles form a planar structure, such that the perpendicular to this plane is aligned with the CMB dipole. 

\subsubsection{Alignment of galaxy axes and polarizations:} 
There have been many observations, both in optical \citep{Hutsemekers:1998} and radio \citep{Tiwari:2012rr,Taylor_align} data sets that suggest alignment of galaxy axes and integrated linear polarizations. These observations can be nicely explained in terms of the correlated magnetic field which may be of primordial origin \citep{Tiwari:2015si}. Intriguingly, the optical alignment is seen to be very prominent in the direction of the CMB dipole \citep{Ralston:2004}.

\subsubsection{Dipole in radio polarization offset angles:} The integrated polarizations of radio galaxies are known to be aligned approximately perpendicular to the galaxy position axes. Remarkably, the angle between these two axes shows a dipole pattern in the sky with preferred axis again pointing roughly along the CMB dipole \citep{Jain:1998r}. Hence, we see that several diverse observations appear to indicate the same preferred direction. Taken together, they are strongly suggestive of a violation of the CP \citep{Ralston:2004}.

\subsubsection{Dipole modulation and the Hemispherical Asymmetry:} We find that the CMB temperature fluctuations have slightly higher power in the southern ecliptic hemisphere than the northern one. This is called the \textit{hemispherical power asymmetry} and was first observed in the WMAP data \citep{Hoftuft2009} and continues to persist in the Planck measurements \citep{Planck:2020_results2018}. It is also observed that the CMB temperature fluctuations appear to be modulated by a dipole that points close to the south ecliptic pole. This implies that the CMB temperature fluctuation along line-of-sight direction $\mathbf{\hat{n}}$ is given by:
\begin{equation}
    \Delta T(\hat n) = \Delta T_{\rm iso}\left[1 + A \bm{\lambda} \cdot \mathbf{\hat{n}} \right],
\end{equation}
where $\Delta T_{\rm iso}$ satisfies CP,  $A$ is the amplitude of the  dipole and $\bm{\lambda}$ is the preferred direction. Current Planck measurements \citep{Planck:2020_results2018} give $A=0.070^{+0.032}_{-0.015}$ and $\bm{\lambda} = (221^\circ, -21^\circ) \pm 31^\circ$. Such a dipole modulation would lead to difference in powers in the two hemispheres along $\hat{\lambda}$. 

\subsubsection{Other CMB observations:}
Other observations of SI violations in the CMB are low in significance, albeit they are present in both WMAP and Planck data. For low-$\ell$ values, the even multipoles are anomalously smaller than the odd multipole modes in power. This is called the \textit{parity asymmetry}. The largest asymmetry are evidenced in the lowest multipoles, viz., $\ell\in[2,7]$. These low multipoles show an anomalously small power, which is called the \textit{low power on large scales} in the CMB temperature fluctuations. 

\subsection{Hubble Tension}
The Hubble tension is the disagreement in measured value of the Hubble parameter $H_0$ from different methods. The local universe measurements of $H_0$ using the `distance ladder' method with Cepheids and supernovae type Ia (SNIa) or strong lensing systems differ from the measurements from the CMB assuming $\Lambda$CDM. Other methods like tip of the red giant branch (TRGB) \citep{Reid:2019ApJ} or gravitational wave events \citep{Gayathri:2020GW,Mukherjee:2020GW} have measured value somewhere between the two. Broadly speaking, $H_0$ measurements from the local universe is larger than the measurements from the CMB at nearly $5\sigma$ significance \citep{Anchordoqui:2019yzc}. It has been suggested that the Hubble tension may lead to a breakdown of FLRW metric based cosmology (see \cite{Krishnan:2021dyb} for more details).

The Cepheid-SNIa measurements use Cepheid variables in host galaxies of SNIa, to calibrate the distance. These calibrated type Ia supernovae are then used to calibrate magnitude and redshift of a large sample of SNIa. The full sample of SNIa probes the Hubble flow and is used to directly infer the Hubble parameter. \citet{Riess:2019} estimate the value $H_0 = 74.03 \pm 1.42$ ${\rm km s}^{-1}{ \rm Mpc}^{-1}$. This agrees with the \citet{Freedman:2012} estimate of $H_0= 74.3 \pm 2.1$ ${\rm km s}^{-1}{ \rm Mpc}^{-1}$. The H0LiCOW team's \citep{Wong:2020MNRAS} recent measurement, using the time delay for a system of six gravitationally lensed quasars, yields $H_0=73.3^{+1.7}_{-1.8}$ ${\rm km s}^{-1}{ \rm Mpc}^{-1}$ that agrees very well with Cepheid \citep{Freedman:2001} measurements.

{In addition to the aforementioned `direct' measurements}, CMB can also be used to infer the value of the Hubble parameter `indirectly'. The CMB $T$ and $E$ mode measurements are used to fit the $\Lambda$CDM model. In its basic form, $\Lambda$CDM has only six parameters. The Hubble parameter can be estimated indirectly from the best fit. This indirect estimation of the $H_0$ gives a value lower than the direct measurements. \citet{Planck_results:2018} gives $H_0=67.27 \pm 0.60$ ${\rm km s}^{-1}{ \rm Mpc}^{-1}$ using only  $T$ and $E$ mode data. Estimates of $H_0$ using other CMB experiments like ACT \citep{Dunkley:2010ge} and SPTpol (for $\ell<1000$) \citep{SPT:2017jdf}  give consistent results with Planck. 

\section{Superhorizon perturbation model}
It has been suggested that the superhorizon perturbations can explain the observed violations of statistical isotropy. 
These are perturbations with wavelengths larger than the particle horizon \citep{Erickcek:2008a}. Such modes necessarily exist in a cosmological model. However, in order to explain the the observed violations of isotropy \citep{Gordon:2005} we also need them to be aligned with one another. In \cite{Gordon:2005}, such an alignment is attributed to a stochastic phenomenon known as \textit{spontaneous breakdown of isotropy}. {Alternatively, the alignment may be attributed to an intrinsic violation of the cosmological principle. A very interesting possibility is presented in \cite{Aluri:2012Pre-Inf} and \cite{Rath:2013}.  It is argued that during its very early phase, the Universe may not be isotropic and homogeneous. As explained in {\S\ref{sec:Intro}}, it acquires this property during inflation \citep{Wald:1983}. The modes which originate during the early phase of inflation when the Universe deviates from isotropy and homogeneity may not obey the cosmological principle \citep{Rath:2013}. We postulate that these are the aligned superhorizon modes.  }

\subsection{Resolution of various anomalies}
Cosmological implications of this phenomenon have been obtained by assuming the existence of a single {adiabatic} mode \citep{Erickcek:2008a,Erickcek:2008b,Shamik:2014,Das:2021JCAP,Tiwari:2022HT}.
Working in the conformal Newtonian Gauge, 
such a mode can be expressed as,
\begin{equation}
    \Psi_\mathrm{p}=\varrho\sin(\kappa x_3 + \omega)\label{eq:SuperMode}
\end{equation}
{Thus a superhorizon mode is characterised by its amplitude $\varrho$, wavenumber $\kappa$ and phase factor $\omega\ne 0$.}
In Eq. \eqref{eq:SuperMode}, we have taken the mode to be aligned along the $x_3$ (or $z$) axis which we also assume to be the direction of CMB dipole. {For a superhorizon mode, we have $\kappa / H_0 \ll 1$.}

It has been shown that such a superhorizon mode is consistent with all existing cosmological observations (like CMB, NVSS constraints etc.) for a range of parameters \citep{Shamik:2014,Das:2021JCAP,Tiwari:2022HT}. Some parameter values are given in Table \ref{tab:SuperParams}. It
can affect the large scale distribution of matter and can potentially explain the enigmatic excess  dipole signal observed in the radio galaxy distribution \citep{Singal:2011,Gibelyou:2012,Rubart:2013, Tiwari:2014ni,Tiwari:2015np,Tiwari:2016adi,Colin:2017}.

{The observed matter dipole, $D_\mathrm{obs}$} {is expressed as:
\begin{equation}
 \mathbf{D}_\mathrm{obs} = \big( D_\mathrm{kin} +  D_\mathrm{grav} +  D_\mathrm{int} \big)\hat{x}_3,
\label{eq:super_dipole}
\end{equation}
where $ D_\mathrm{kin},\  D_\mathrm{grav}$ and  $D_\mathrm{int}$ respectively denote the amplitudes of the kinematic, gravitational and intrinsic dipoles. These components are redshift dependent. Thus we can write the magnitude of the observed dipole between the redshifts $z_1$ and $z_2$,} {due to the superhorizon mode \eqref{eq:SuperMode}} {as
\begin{align}
D_\mathrm{obs}(z_1,z_2) &= \Big[\mathcal{A}_1(z_1, z_2) + \mathcal{A}_2(z_1, z_2)\notag \\
&+ \mathcal{C}(z_1, z_2)\Big]\frac{\varrho\kappa\cos\omega}{H_0}+ \mathcal{B} \label{eq:DipoleZDepend}
\end{align}
where the term 
\begin{equation}
    \mathcal{B} = [2+x(1+\alpha)]\frac{v}{c}\label{eq:KineDipole}
\end{equation}
is the redshift independent kinematic dipole component. The explicit expressions for other redshift dependent factors $\mathcal{A}_1(z_1, z_2)$,  $\mathcal{A}_2(z_1, z_2)$, $\mathcal{C}(z_1, z_2)$ are given in \citep{Das:2021JCAP}. Notice that in the absence of a superhorizon mode, i.e., $\varrho\to0$, the dipole magnitude in Eq. \eqref{eq:DipoleZDepend}, as expected, becomes redshift independent and equal to \eqref{eq:DipNoSuperHori}.}

\subsubsection{The Matter Dipole:}
{
{Due to the presence of an aligned superhorizon mode, an additional contribution to our velocity arises with respect to LSS in the CMB dipole direction. This is given in Eq. (2.12) of \citep{Das:2021JCAP}}. Hence it leads to a change in $D_\mathrm{kin}$ in comparison to its prediction based on CMB dipole \citep{Das:2021JCAP}. Furthermore, the superhorizon mode contributes through the Sachs-Wolfe (SW) and the integrated Sachs-Wolfe (ISW) effects \citep{Erickcek:2008a}, thereby leading to $D_\mathrm{grav}$ in Eq. \eqref{eq:super_dipole}. Finally, the superhorizon mode leads to an intrinsic anisotropy in the matter distribution and hence contributes to $D_\mathrm{int}$. Eq. \eqref{eq:DipoleZDepend} is the explicit expression considering all these effects. From the equation, it is clear that for a given value of $\varrho>0$, the dipole contribution is maximum if $\omega=\pi$. All the contributions due to the mode depend on redshift since the mode has a systematic dependence on distance and hence the predicted dipole is redshift dependent.}

{It is interesting to note that the contributions of the superhorizon mode to the CMB dipole cancel out at the leading order \citep{Erickcek:2008a}. Such a cancellation does not happen in the case of matter dipole \citep{Das:2021JCAP}. We may understand this as follows. As per Eq. \eqref{eq:super_dipole}, there are three different contributions to the matter dipole -- (a) the kinematic dipole which arises due to our velocity relative to the source, (b) the gravitational dipole (SW and ISW) and (c) the intrinsic dipole. In the case of CMB, these three add up to zero. In the case of matter dipole, the kinematic dipole explicitly depends on the parameter $\alpha$ which arises in the spectral dependence of the flux from a source, as well as the parameter $x$ (see Eq. \eqref{eq:KineDipole}) which arises in the number count distribution. Furthermore, the gravitational effect also depends on $\alpha$. The intrinsic dipole, however, does not depend on either of these parameters. We point out that both of these parameters arise at non-linear order in {the theory of} structure formation and furthermore the assumed power law distribution is only an approximation \citep{Tiwari:2014ni}. These parameters are best extracted from observations and cannot be reliably deduced theoretically. Hence, the situation is very different in the case of matter dipole in comparison to CMB dipole and we do not expect that the two would behave in the same manner. We clarify that in the case of matter dipole, the superhorizon {perturbation} is treated at first order in perturbation theory. However, the small wavelength modes which are responsible for structure formation have to be treated at nonlinear order. In \cite{Das:2021JCAP}, the existence of structures is assumed as given with their properties deduced observationally and the calculation focuses only on the additional contribution due to the superhorizon mode. However, a complete first principles calculation would have to treat small wavelength modes at nonlinear order. }

{
There are some further issues (see \cite{Domenech:2022mvt} for more details), associated with gauge invariance \citep{Challinor_2011,Bonvin_2011}, which are not addressed in \cite{Das:2021JCAP}. These issues are very important, but to the best of our understanding, they are expected to lead to small corrections to the calculational framework used in \cite{Das:2021JCAP} and not expected to qualitatively change their results. It will be very interesting to repeat these calculations using the gauge invariant framework, but this is beyond the scope of the present paper. Such a calculation must also take into account the fact that the aligned superhorizon modes we are considering do not arise within the $\Lambda$CDM model but perhaps due to an anisotropic/inhomogeneous early phase of cosmic expansion \citep{Aluri:2012Pre-Inf,Rath:2013}.}

\subsubsection{Alignment of Quadrupole and Octupole:}
{Further, the superhorizon mode can also explain the alignment of CMB quadrupole and octupole \citep{Gordon:2005ai}. With $x_3$ axis along the CMB dipole, it leads to non-zero spherical harmonic coefficients $T_{10}$, $T_{20}$, $T_{30}$ in the temperature anisotropy field \citep{Erickcek:2008a}. We obtain  constraints on the mode parameters in Eq. \eqref{eq:SuperMode} by requiring that 
$T_{20}$ and $T_{30}$ are less than three times the measured rms values of the quadrupole and octupole {powers} respectively \citep{Erickcek:2008a}. It turns out that the dipole contribution does not lead to a significant constraint. These contributions can explain the alignment of quadrupole and octupole if we assume the presence of an intrinsic contribution to $T_{10}$ and $T_{20}$ which is partially cancelled by the contribution due to the superhorizon mode. Note that this intrinsic contribution is statistical in nature and hence its exact value cannot be predicted.}
\subsubsection{Hubble Tension:}
{It has been shown by \cite{Das:2021JCAP} that a superhorizon mode leads to a perturbation in the gravitational potential between distant galaxies and us. This culminates in a correction in observed redshift of galaxies.
\begin{equation}
    1+z_\mathrm{obs} = (1+ z)(1+ z_{\rm Doppler})(1+ z_{\rm grav})
\end{equation}
Thus we see that in the presence of superhorizon modes, the galaxy at redshift $z$ is observed instead at a redshift $z_\mathrm{obs}$. In the above equation, the redshifts  $z_\mathrm{Doppler}$ and $z_\mathrm{grav}$ are respectively  due to our velocity relative to LSS and perturbation in potential introduced by the superhorizon mode. {We can express $z_\mathrm{obs}$ as \citep{Das:2021JCAP,Tiwari:2022HT},
\begin{equation}
    z_\mathrm{obs} = \bar z + \gamma \cos\theta + \ldots \label{eq:zObsExpan}
\end{equation}
where the first and the second terms on the RHS are the monopole and dipole terms. Here $\theta$ is the polar angle of the source with $x_3$ axis along the CMB dipole and $\gamma$ the dipole amplitude. Interestingly, the monopole term in Eq. \eqref{eq:zObsExpan} resolves the Hubble tension \citep{Tiwari:2022HT}. For that we need to choose the phase $\omega\ne \pi$. The range of parameters which explain both the matter dipole and the Hubble tension is given in \citep{Tiwari:2022HT}. In Table \ref{tab:SuperParams}, we quote some of those values.} 
}

The superhorizon modes are also likely to  leave their signatures in other cosmological observables like Baryon Acoustic Oscillations, epoch of reionziation etc. 
{
\begin{table}
\centering
\begin{tabular}{ p{.5cm}p{2.0cm}p{2.0cm}p{2.0cm}}
	\hline
	\hline
	\noalign{\vskip 0.1cm}
	  & $\omega$ & $\varrho$ & $\kappa/H_0$\\
	\hline
	\noalign{\vskip 0.1cm}
	1 & $0.81\pi$ & $0.97$ & $2.58\times 10^{-3}$ \\
	2 & $0.81\pi$ & $0.48$ & $6.4\times 10^{-3}$ \\
	\hline
	\hline
\end{tabular}
\caption{Some parameter values for the superhorizon mode (Eq. \ref{eq:SuperMode}) explaining NVSS excess dipole and also resolving the Hubble tension. These values also satisfying CMB constraints are taken from \cite{Tiwari:2022HT}.}
\label{tab:SuperParams}
\end{table}}

\section{Constraints using SKA}

\subsection{Superhorizon perturbation observation}

The superhorizon model predicts several observations which can be tested with SKA and other future surveys. An interesting feature is the significant dependence of dipole on the redshift in the presence of superhorizon modes. Hence, the dipole measurements in redshift bins with SKA1 and SKA2 continuum survey can work as a potential test of the model. {For a radio continuum survey, it is unlikely that we would have spectroscopic redshift information. In the past, redshifts of radio galaxies have been estimated by taking cross correlation with well known redshift surveys \citep{Blake:2002, Tiwari:2014ni}. Such strategies are still viable by using data from the GAMA survey fields \citep{Baldry:2018}. However, new techniques like template fitting \citep{Duncan:2018} or machine learning based photometric redshift computations \citep{Brescia:2021} make it possible for the SKA radio continuum survey galaxies to contain redshift information. This added information provides a unique possibility to test superhorizon mode physics with the SKA.}  

\subsection{Predictions}
Here, we demonstrate how precisely evident the dipole predictions with a superhorizon model would be. Further, we demonstrate how much they are constrained using SKA observations. Assuming superhorizon modes that (a) satisfy the present NVSS and Hubble parameter measurements  and (b) are consistent with CMB and other cosmological measurements (see Table \ref{tab:SuperParams}); we obtain the dipole magnitude $D_\mathrm{obs}$ in redshift bins using the formalism described in \citep{Tiwari:2022HT}. The dependence of dipole signal {on the redshift $z$ is shown in Figure \ref{fig:dipole} where we have shown the redshift dependence of $D_\mathrm{obs}$ for two cases
\begin{enumerate}
\item {Cumulative Redshift Bins:} For this case, we fix $z_1=0$ in Eq. \eqref{eq:DipoleZDepend} and vary $z_2=z$. In other words, we calculate $D_\mathrm{obs}(0,z)$.  
\item {Non-overlapping Redshift Bins:} In this case, we obtain the non overlapping $z$ dependence by evaluating $D_\mathrm{obs}(z-\Delta z,z+\Delta z)$ with $\Delta z= 0.25$. For a fix $\Delta z$, this thus gives $D_\mathrm{obs}$ at $z$.
\end{enumerate}
\begin{figure}
    \centering
    \includegraphics[scale=0.28]{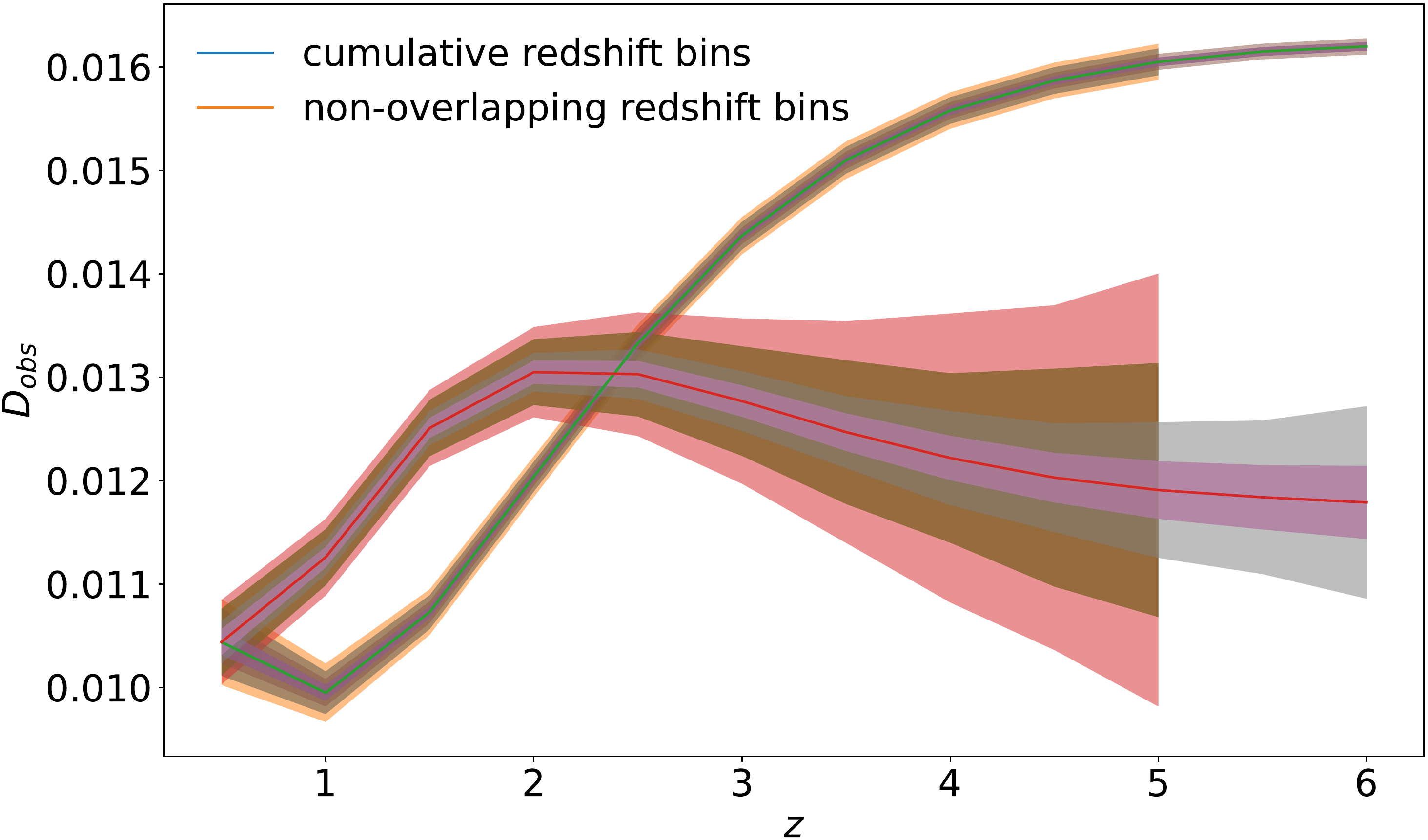}
\caption{The dipole signal observation in the presence of superhorizon modes with SKA1 {($z\le 5$)} and SKA2 {($z\le 6$)} continuum surveys. {The inner and outer shaded regions respectively represent the optimistic and realistic uncertainties for both SKA1 \& SKA2.} 
Here we have assumed a superhorizon mode \citep{Tiwari:2022HT} satisfying present NVSS dipole observation \citep{Tiwari:2014ni}.}
    \label{fig:dipole}
\end{figure}

\subsection{Estimating Uncertainties}}
We employ \cite{Alonso:2015} `Ultra-large scales'  codes\footnote{\href{http://intensitymapping.physics.ox.ac.uk/codes.html}{http://intensitymapping.physics.ox.ac.uk/codes.html}} (for continuum surveys) to determine the number densities for SKA surveys. We further assume that SKA1 and SKA2 will observe the sky up to respective declinations of $15^\circ$ and $30^\circ$. The optimistic and realistic flux densities' limits for SKA1 and SKA2 \citep{Bacon:2018,Bengaly:2018} are given in Table \ref{tab:RealOpt}.
{
\begin{table}
    \centering
    \begin{tabular}{p{2.5cm}p{1.5cm}p{1.5cm}}
    \hline
    \hline
    \noalign{\vskip 0.1cm}
    Flux Density & SKA1 & SKA2 \\
    \hline
    \noalign{\vskip 0.1cm}
    Optimistic & $>10$ & $>1$ \\
    Realistic & $>20$ & $>5$ \\
    \hline
    \hline
    \noalign{\vskip 0.1cm}
    \end{tabular}
    \caption{Optimistic and realistic flux densities (in $\mu$Jy) for SKA1 and SKA2 surveys.}
    \label{tab:RealOpt}
\end{table}}

Additionally, we note that SKA1 is expected to probe up to $0\le z\le 5$, whereas the SKA2 will reach up to redshift 6. We mock SKA1 and SKA2 continuum sky to determine the  observational implications of the superhorizon model. We produce 1000 number density simulation of SKA1 and SKA2 continuum survey for each (optimistic and realistic) flux threshold using \textsc{HEALPix} software \citep{Gorski:2005}, with $N_{\rm side}=64$. The mean number of galaxies in a pixel is determined using number density obtained from \cite{Alonso:2015} code and by modelling a dipole 
with magnitude and direction expected in presence of a superhorizon mode. Given the mean number density in a pixel, we call random Poisson distribution to emulate the galaxy count in the pixel. The galaxy mock thus neglects the cosmological galaxy clustering.  This is justified since {the clustering dipole in LSS is $\approx 2.7 \times 10^{-3}$ \citep{Adi:2015nb,Tiwari:2016adi}, which is roughly five times less than the apparent dipole in LSS\footnote{These estimates correspond to NVSS galaxies. Assuming the NVSS measured dipole in LSS is true, we expect similar numbers from SKA continuum surveys.} and inconsequential for our simulations. This is 
roughly equal to the uncertainties in the measured dipole is LSS using NVSS galaxies (see Table \ref{tab:lss_dipole}).} We consider SKA1, SKA2 sky coverage, i.e., mask declination {above}  $15^\circ$ and $30^\circ$, respectively.  Additionally, we mask the galactic latitudes ($|b| < 10^\circ$) in order to remove Milky Way contamination. {The galactic plane cut is often chosen to be anywhere between $|b| < 5^\circ$ and $|b| < 15^\circ$, and in most studies one tests the robustness of the results with varying redshift cuts. For tests on mock data, here we choose a typical cut of $|b| < 10^\circ$ that should balance the exclusion of galactic plane contamination and loss of sky fraction.} Next, we use \texttt{Python Healpy}\footnote{\href{https://healpy.readthedocs.io/en/latest/index.html}{https://healpy.readthedocs.io/en/latest/index.html}} \citep{Gorski:2005,Zonca2019} \texttt{fit\_dipole} function and obtain dipole for each 1000 mock maps. From these 1000 dipole values, we calculate the standard deviation to determine the uncertainty in measurements. The shaded regions in Figure \ref{fig:dipole} show the results obtained for SKA1 and SKA2 optimistic and realistic number densities in non-overlapping and cumulative redshift bins.

\subsection{Other Anisotropy tests with SKA}
If the universe does not follow CP at large distance scales then every observable should have directional dependence characteristics. Out of all, three observables are of particular interest as these are independent of the number density over the sky. So these are more robust under unequal coverage and systematics of the sky. These observables are

\begin{itemize}
\item {Mean spectral index $(\bar{\alpha})$ -- As we said in \S\ref{sec:KineDipo}, the spectral index for a radio source is defined between flux density and frequency through $S \propto \nu^{-\alpha}$. In the \texttt{Healpy} pixelation scheme, the sky is divided into equal area pixels. For a given pixel p with $N_\mathrm{p}$ sources having spectral indices $\alpha_{i,\mathrm{p}}$, we define mean spectral index
	\begin{equation}
 \bar{\alpha}_{\mathrm{p}} = \frac{1}{N_\mathrm{p}}\sum_i\alpha_{i,\mathrm{p}}
	\end{equation}
here $i$ runs over all the sources in pixel p}
	
\item Exponent $(x)$ of differential number count -- Defined using $N(S, \mathbf{\hat {n}}) \propto S^{-1-x}$
	
\item {Average Flux Density $(\bar{S})$ -- We define this quantity for a pixel p
\begin{equation}
    \bar{S}_\mathrm{p} = \frac{1}{N_\mathrm{p}}\sum_i S_{i,\mathrm{p}}
\end{equation}
where again $i$ runs over all the sources in p and $N_\mathrm{p}$ is the number of sources in the pixel}
\end{itemize}

The spectral index characterises the morphology of an astronomical source. Angular dependence of $\bar{\alpha}$ has not been much looked at  in the literature. Analysing the dipole anisotropy in $\bar{\alpha}$ has been a challenge since it requires reliable multi-frequency continuum radio sky survey. Such an SKA survey can be used to estimate this anisotropy if the flux density of the sources at different frequencies is measured with sufficient accuracy. For $x$, the angular dependency was analysed by \citep{Ghosh_2017} in NVSS data using likelihood maximisation and the results were found to be consistent with CP. However, with a larger expected source count of SKA that is almost twice in comparison to NVSS, angular dependence analysis of $x$ may provide a more stringent test of CP.

SKA will also be able to test the phenomenon of alignment of radio galaxy axes and integrated polarizations, as claimed in earlier radio observations \citep{Tiwari:2012rr,Tiwari:2015si,Taylor_align}. 

\section{Conclusion and Outlook}
In this paper, we have reviewed several cosmological signals which appear to show a violation of CP. We have also reviewed a model, based on aligned superhorizon modes which can explain some of these observations along with the Hubble tension \citep{Tiwari:2022HT}. The model can be theoretically justified by postulating a pre-inflationary phase during which the Universe may not be homogeneous and isotropic \citep{Rath:2013}. The model leads to several cosmological predictions which can be tested at SKA. By using the best fit parameters with current observations, we have determined the redshift dependence of the predicted dipole in radio galaxy number counts and associated uncertainties. As can be seen from Figure \ref{fig:dipole}, SKA can test this very reliably. If this prediction is confirmed by SKA, it may provide us with a first glimpse into the Physics of the pre-inflationary phase of the Universe.

{We have also suggested other isotropy tests with SKA using other variables which are independent of number density and thus are more robust under unequal coverage and systematics of the sky. These variables are (a) mean spectral index $\bar{\alpha}$, (b) exponent of the differential number count $x$ \& (c) average flux density $\bar{S}$.}

\section*{Acknowledgements}
RK is supported by the South African Radio Astronomy Observatory and the National Research Foundation (Grant No. 75415). PT acknowledges the support of the RFIS grant (No. 12150410322) by the National Natural Science Foundation of China (NSFC) and the support by the National Key Basic Research and Development Program of China (No. 2018YFA0404503) and NSFC Grants 11925303 and 11720101004. SG is supported in part by the National Key R \& D Program of China (2021YFC2203100), by the Fundamental Research Funds for the Central Universities under grant no: WK2030000036, and the NSFC grant no: 11903030. {We are also very thankful to the anonymous referee whose comments were really helpful in improving the presentation of the paper.}

\bibliographystyle{apj}
\bibliography{master_Hubble,master_radio}

\end{document}